\documentclass[aps,prb,twocolumn,groupedaddress,showpacs]{revtex4} 
\usepackage{graphicx} 

\begin{document}  
  
\title{High pressure ionic and molecular crystals of 
  ammonia monohydrate within density functional theory}

\author{Gareth I. G. Griffiths}  
  
\affiliation {Theory of Condensed Matter Group, Cavendish Laboratory,
  J J Thomson Avenue, Cambridge CB3 0HE, United Kingdom}

\author{Alston J. Misquitta}  
  
\affiliation {Theory of Condensed Matter Group, Cavendish Laboratory,
  J J Thomson Avenue, Cambridge CB3 0HE, United Kingdom}

\author{A. Dominic Fortes} 
 
\affiliation {Centre for Planetary Sciences, University College
  London, Gower Street, London WC1E 6BT, United Kingdom}
 
\author{Chris J. Pickard} 
 
\affiliation {Department of Physics and Astronomy, University College
  London, Gower Street, London WC1E 6BT, United Kingdom}

\author{Richard J. Needs}  
  
\affiliation {Theory of Condensed Matter Group, Cavendish Laboratory,
  J J Thomson Avenue, Cambridge CB3 0HE, United Kingdom}

\date{\today}

\begin{abstract}
  The following article has been submitted to The Journal of Chemical
  Physics. After it is published, it will be found at
  \url{http://jcp.aip.org/}

  A combination of first-principles density functional theory
  calculations and a search over structures predicts the stability of
  a proton-transfer modification of ammonia monohydrate with space
  group $P4/nmm$. The phase diagram is calculated with the PBE density
  functional, and the effects of a semi-empirical dispersion
  correction, zero point motion, and finite temperature are
  investigated. Comparison with MP2 and coupled cluster calculations
  shows that the PBE functional over-stabilizes proton transfer phases
  because too much electronic charge moves with the proton.  This
  over-binding is partially corrected by using the PBE0 hybrid
  exchange-correlation functional, which increases the enthalpy of
  $P4/nmm$ by about 0.6 eV per formula unit relative to phase I of
  ammonia monohydrate (AMH-I) and shifts the transition to the proton
  transfer phase from the PBE pressure of 2.8 GPa to about 10 GPa.
  This is consistent with experiment as proton transfer phases have
  not been observed at pressures up to $\sim$9 GPa, while higher
  pressures have not yet been explored experimentally.
\end{abstract}


\maketitle 
\section{Introduction}

Ammonia monohydrate (AMH, NH$_3$H$_2$O) exists as at least six
different crystalline polymorphs over the experimentally studied range
of pressures and temperatures of 0 $< p <$ 9 GPa and 170 $< T <$ 295
K.\cite{loveday2004ammonia} The crystal structures of three of these
polymorphs have been determined: the low pressure phase
AMH-I,\cite{olovsson1959crystal,nelmes_airapt} the high pressure
disordered body-centred-cubic phase AMH-VI\cite{loveday1999ammonia}
and, most recently, a combination of \emph{ab initio} random structure
searching (AIRSS)\cite{airss_review_2011,pickard:045504} and neutron
powder diffraction data led to the solution of the crystal structure
of AMH-II.\cite{fortes2009crystal,10.1063/1.3245858} The structures of
AMH-III, IV and V remain to be determined.  The crystal structures and
properties of AMH polymorphs (and of the related compound ammonia
dihydrate, ADH) are of interest to planetary scientists due to the
likely presence of substantial fractions of ammonia in ice accreted
into the satellites of the Gas Giant
planets.\cite{kargel1992ammonia,springerlink:10.1007/s11214-010-9633-3}
Whilst the water-rich compound (ADH) may have a greater abundance than
AMH at low pressures, it is known that a high-pressure form of ADH
becomes unstable with respect to a mixture of high-pressure AMH and
water ice at $\sim$3.5 GPa.\cite{fortes2009phase} Such a pressure is
relevant to the core of the large icy satellite Titan if it
undifferentiated (of uniform composition),\cite{Fortes2011}
as well as during the period of its
accretion.\cite{lunine1987clathrate}
Such a pressure may also occur in the icy mantles of fully
differentiated giant icy exoplanets or exomoons.\cite{fu2010interior}
NH$_3$, H$_2$O and CH$_4$ are likely to comprise a substantial
fraction of the interiors of Uranus and Neptune at pressures up to 600
GPa and temperatures up to 7000 K.\cite{hubbard1980structure} The
properties of this high $p$-$T$ molecular mixture are thought to be
important in the generation of unusual magnetic fields in these
bodies.\cite{cavazzoni1999superionic}

Neutron single-crystal diffraction, and the indexing and solution of a
structure from powder diffraction data are non trivial at high
pressure. This, combined with the tendency towards the formation
and/or persistence of metastable phases in low-temperature condensed
molecular systems on laboratory timescales, makes it clear that there
is a central role for the computational prediction of equilibrium
crystal structures and the determination of their physical properties.

In this work we combine first-principles density functional theory
(DFT) calculations with a random search strategy in order to identify
new candidate high-pressure AMH structures and to compute their
stability with respect to one another and to other known crystalline
polymorphs of AMH.  Earlier DFT studies of ammonia monohydrate,
ammonia hemihydrate (AHH) and solid ammonia, have revealed a
propensity towards proton transfer (i.e., formation of an ionic solid)
at high pressures. Calculations have suggested that AMH-I transforms
to ammonium hydroxide at $\sim$5 GPa,\cite{fortes2001ab} AHH
transforms to ammonium hydroxide ammoniate at $\sim$12
GPa,\cite{fortesphd} and solid ammonia transforms to ammonium amide at
$\sim$90 GPa.\cite{pickard2008hca} The ionic solids derived from the
two hydrates are isosymmetric with their molecular precursors, but
there is no reason to suppose that other ionic structures might not be
energetically stable.

The presence of weak hydrogen bonds and the occurrence of both homo-
and hetero-nuclear hydrogen bonds in the ammonia hydrates provide a
challenge for electronic structure methods, particularly with respect
to the accuracy of exchange-correlation functionals. DFT calculations
using standard functionals such as the Perdew-Burke-Ernzerhof (PBE)
generalized gradient approximation (GGA) predict hydrogen-bonded
molecular phases of AMH at low pressures, in agreement with
experiment.  We show here, however, that this approach predicts
molecular AMH phases to be unstable to the formation of ionic ammonium
hydroxide (NH$_4^+$:OH$^-$) proton-transfer phases at pressures of
about 2.8 GPa, although no experimental evidence for such phases has
been found to date, even at pressures up to about 9 GPa.  The weak van
der Waals forces, which are not described by standard density
functionals such as the PBE-GGA, turn out to be important in
determining the volumes and relative enthalpies of the phases in this
system.  The zero-point (ZP) motion of the H atoms is also important
for an accurate account of the energetics.  We have, moreover, found
that the most serious defect of PBE calculations in this system is
that the energetics of the proton transfer is very poorly described as
too much electronic charge is transferred with the proton.  We show
that a satisfactory description of the experimental data, including
the absence of proton-transfer phases at low pressures, requires the
inclusion of nuclear ZP motion and accurate descriptions, beyond those
afforded by functionals such as PBE-GGA, of both exchange interactions
and the van der Waals forces that arise from electron correlation.

\section{\textit{Ab initio} Random structure searching} 

We have used the AIRSS method\cite{airss_review_2011,pickard:045504}
to identify low-enthalpy structures of AMH at pressures of up to about
12 GPa.  In the AIRSS approach randomly chosen structures are relaxed
to a minimum in the enthalpy at fixed pressure.  In its simplest form
AIRSS has almost no free parameters and is essentially unbiased, and
it is therefore the ideal basis upon which to impose constraints and
biases towards the types of structure that one believes are most
favorable.  Perhaps the simplest physical constraint that we have
employed is to reject initial configurations in which atoms are closer
than a defined minimum separation.  One of the most useful constraints
is to restrict the symmetries of the structures.  The structures are
chosen to obey the symmetries of a particular space group, although
they are otherwise random, and the desired symmetry is maintained
throughout the relaxation procedure. Another useful approach is to
choose initial structures constructed from randomly placed ``chemical
units'', which in this case are equal numbers of NH$_3$ and H$_2$O
molecules, or equal numbers of NH$_4$ and OH units, or
``hydrogen-bonded'' NH$_3$$\cdot$H$_2$O AMH units.  Each initial unit
cell was generated by choosing random unit cell translation vectors
and scaling the volume to lie randomly within $\pm$50\% of some
reasonable value.
We then placed the required number of chemical units within the cell,
applying symmetry constraints as required.  Initial structures in
which the overlap of molecules was significant were rejected because
they are likely to undergo unwanted chemical reactions during the
relaxation procedure.

We first performed unconstrained searches with $2$ formula units
(f.u.), and we then searched with $4$ f.u.\ and initial structures
formed from NH$_3$ and H$_2$O molecules.  Searches were then performed
starting from random arrangements of either 2 or 4 preformed AMH
units.  Symmetry constrained searches were performed starting from
random cells containing $2$ randomly placed H$_2$O molecules and $2$
randomly placed NH$_3$ molecules, and then applying the symmetry
operations of space groups randomly chosen from those with 2
operations.  Similar searches were performed with $4$ symmetry
operations and single units of H$_2$O and NH$_3$.  To bias the
procedure towards finding ionic structures, we started further
searches with cells containing $4$ f.u.\ and using building blocks of
NH$_4$ and OH units with $n=2$ or $n=4$ symmetry operations.  $3$ and
$5$ f.u.\ searches were performed without symmetry constraints.  We
also performed searches over larger unit cells containing $6$, $6$ and
$8$ f.u.\ in total, generated with $n=6$, $n=3$ and $n=8$ symmetry
operations, respectively.  A total of about 7,700 structures were
relaxed during the searches.

The {\sc CASTEP} plane wave code\cite{clark2005first} was used for all
of the calculations on periodic crystals.  Calculations were performed
with the Perdew-Burke-Ernzerhof (PBE) generalized gradient
approximation (GGA) exchange-correlation density
functional,\cite{PhysRevLett.77.3865} the PBE functional with the
Grimme semi-empirical dispersion correction
(G06),\cite{grimme2006semiempirical} and the PBE0 hybrid density
functional\cite{adamo1999toward} which includes 25\% exact
(Hartree-Fock) exchange.  We used ultrasoft
pseudopotentials\cite{PhysRevB.41.7892} for the PBE and PBE+G06
calculations and norm-conserving pseudopotentials generated using the
Opium software\cite{website:opium} for the PBE0 calculations.  For the
searches we used a plane wave cut off energy of $340$ eV and a
Monkhorst-Pack\cite{PhysRevB.13.5188} Brillouin zone sampling grid of
spacing $2\pi \times 0.07$ \AA$^{-1}$, and $2\pi \times 0.08$
\AA$^{-1}$ for some of the searches with larger unit cells.  All of
the results reported in this paper were obtained by refining the
structures obtained in the searches at a higher level of accuracy
consisting of a plane wave cut off energy of 700 eV and a Brillouin
zone sampling grid of spacing $2\pi \times 0.04$ \AA$^{-1}$.  The
enthalpy difference between AMH I and the ionic $P4/nmm$ phase of AMH
reported here was changed by less than 0.0002 eV per AMH formula unit
(f.u.) on doubling the cut off energy to 1400 eV, while the enthalpy
change on doubling the number of k-points was even smaller.  The
required force tolerance for a successful geometry optimization in
each search run was $0.05$ eV/{\AA}, which was tightened to $0.005$
eV/{\AA} for the final results reported in this paper. The stress on
the unit cell was converged to better than $0.01$ eV/{\AA}$^3$. The
norm-conserving pseudopotentials required a plane wave cut off energy
of 1000 eV, for which the energy difference between AMH-I and $P4/nmm$
was converged to better than 0.00002 eV per f.u.\ A less dense
Brillouin zone sampling grid spacing of $2\pi \times 0.05$ \AA$^{-1}$
was used for the PBE0 calculations, which gave an energy convergence
of better than 0.03 eV per f.u.

\begin{figure}  
\includegraphics*[width=0.45\textwidth]{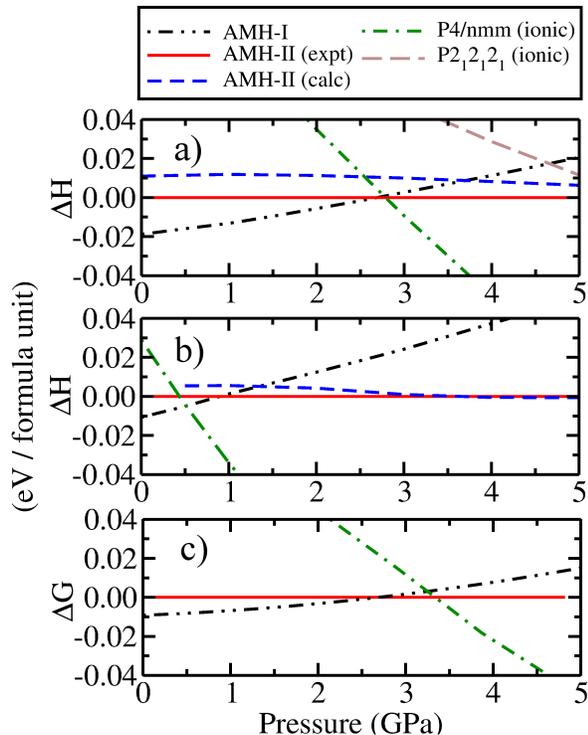}  
\caption{\label{fig:pbe_merged} Enthalpies per f.u.\ of
  AMH structures relative to AMH-II calculated with: a) the PBE
  functional, b) the PBE functional and G06 semi-empirical dispersion
  correction. The Gibbs free energy per f.u.\ of AMH structures
  relative to that of AMH-II is shown in c), calculated with the PBE
  functional including vibrational motion at 175 K.  AMH-II (expt)
  denotes the structure found in experiment while AMH-II (calc)
  denotes the structure found in the AIRSS study reported in Ref.\
  \cite{fortes2009crystal}.}
\end{figure} 

\begin{figure}  
\includegraphics*[width=0.42\textwidth]{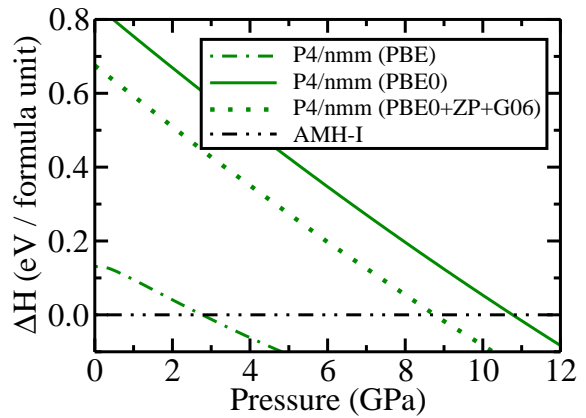}  
\caption{\label{fig:pbe0_merged} Enthalpies per f.u.\
  of $P4/nmm$ relative to AMH-I for three methods; using the PBE
  functional, the PBE0 hybrid functional, and PBE0 with both the G06
  dispersion correction and ZP vibrational motion effects included.
  The PBE0 calculations were performed on the structures relaxed using
  the PBE functional. }
\end{figure}

\section{Results from structure searching} 


 \begin{figure}  
\includegraphics*[width=0.40\textwidth]{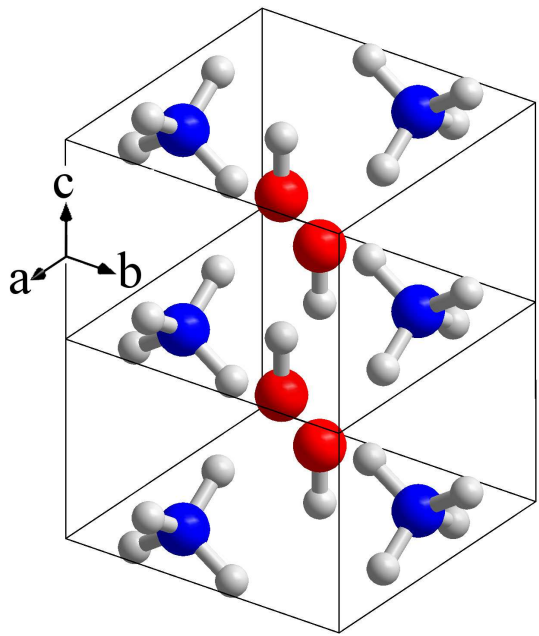}  
\caption{\label{fig:ionic_struct} The $P4/nmm$ ionic
  structure; light grey atoms are hydrogen, red atoms are oxygen, and
  blue atoms are nitrogen.  The solid lines depict two unit-cells, the
  orientation of the crystallographic axes being indicated on the
  left.  }
\end{figure}

AIRSS was used successfully to determine the crystal structure of
AMH-II in collaboration with experiment, which provided initial
constraints on the symmetry and dimensions of the unit
cell.\cite{fortes2009crystal} The AIRSS structure with 112 atoms in
the primitive unit cell was found to be almost correct in a subsequent
experiment, with the exception that it produced one of the two
possible H-bond ordering schemes that is apparently not adopted by the
real material.  The relative enthalpies of the AMH-II structure
obtained from AIRSS and the experimental structure, both fully relaxed
within PBE, are shown in Fig.\ \ref{fig:pbe_merged}.  The PBE and
PBE+G06 calculations indicate that the AMH-II structure obtained from
AIRSS is roughly 0.01 eV higher in enthalpy than the experimental
structure.  There is no reason why the experimental structure with the
alternate H-bond ordering could not have been found if more searches
had been performed.

Fig.\ \ref{fig:pbe_merged}(a) shows the variation of the enthalpy with
pressure for a number of AMH phases calculated with the PBE
functional.  Our searches did not find any molecular structures more
stable than AMH-I and II.  This may suggest that, if any of the three
unsolved AMH polymorphs are molecular, they are likely to have complex
architectures with $Z^{\prime}>8$, where $Z^{\prime}$ is the number of
molecules in the asymmetric unit cell.  The ionic ammonium hydroxide
structure (space group $P2_12_12_1$) was obtained in the DFT study of
Fortes \textit{et al.}\cite{fortes2001ab} by compressing AMH-I which
underwent an isosymmetric transition to the ionic (proton-transfer)
form manifested by a discontinuity in the slope of the calculated
energy-volume curve.  This transition was also observed in the present
work. The ionic $P2_12_12_1$ structure becomes more favourable than
either AMH-I or AMH-II above about $6$ GPa.  Although the $P2_12_12_1$
ionic phase has a region of stability on this phase diagram relative
to the known phases, there is no reason to believe that it is the
\emph{most} stable ionic phase of AMH, which motivates a systematic
search, as described above. Searching with AIRSS revealed a structure
(shown in Fig.\ \ref{fig:ionic_struct}) with space group $P4/nmm$ to
be the most stable in all $2$ f.u.\ searches at $10$ GPa.
Subsequent searches with $4$ f.u.\ at both $3$ and $10$ GPa also
showed $P4/nmm$ to be the most stable structure, regardless of the
constraints imposed. Even searches with $8$ f.u.\ found $P4/nmm$ to be
the most stable.  None of the searches performed with $3$, $5$, or $6$
f.u.\ resulted in structures with enthalpies as low as that of
$P4/nmm$.  In fact none of the space groups with $3$ symmetry
operations are subgroups of $P4/nmm$, and therefore it could not have
been found in searches with $3$ symmetry operations.


The stability range of AMH-II between the AMH-I and ionic $P4/nmm$
phases predicted by the PBE functional is very small ($\sim$0.01 GPa,
see Fig.\ \ref{fig:pbe_merged}).  However, experiments have shown that
the transition from AMH-I to AMH-II takes place at around 0.5 GPa, and
then from AMH-II to AMH-IV (whose structure is presently unknown) at
around 2.2 GPa at 170 K.\cite{loveday2004ammonia,fortes2009crystal}
An experimental study has shown that warming AMH-IV at 6.5 GPa
produces the body-centred-cubic (bcc) phase VI at $\sim$280
K,\cite{loveday2004ammonia} the structure of which has been reported
to consist of orientationally and positionally disordered NH$_3$ and
H$_2$O molecules.\cite{loveday1999ammonia} Our neutron powder
diffraction study has shown that compression of AMH-V at room
temperature does not lead to the formation of AMH-VI up to $\sim$9
GPa.\cite{fortes2010report} Interestingly, a similar disordered bcc
phase of ADH was reported by Fortes \textit{et
  al.}\cite{doi:10.1080/08957950701265029} and confirmed recently by
Loveday \textit{et al.}\cite{doi:10.1080/08957950903162057} It is
likely that a solid solution could exist between the AMH and ADH
compositions at high pressures, if the bcc crystal structure is
maintained over a range of occupancies of the NH$_3$ and H$_2$O
molecules.

Clearly, construction of a complete computational phase diagram
requires simulation of the AMH-VI structure. However, the disordered
nature of AMH-VI precludes straightforward investigation using DFT
but, as a first approximation, a so-called ``shaking'' search was
performed, in which a 2$\times$2$\times$2 supercell comprising only
the oxygen and nitrogen atoms was created. For each search the
appropriate number of hydrogen atoms were distributed randomly over
the supercell and the atomic positions were relaxed. The lowest
enthalpy structure, which was obtained repeatedly, does not appear in
Fig.\ \ref{fig:pbe_merged} as it lies approximately 0.2 eV per f.u.\
above AMH-II and is quite far from thermodynamic stability. The large
discrepancy between the computed stability and the reproducible
experimental observation of AMH-VI is likely to be due to the small
cell used in the calculations.

PBE calculations predict the ionic $P4/nmm$ phase to be
thermodynamically stable across a broad region of the high pressure
phase diagram.  Whether $P4/nmm$ corresponds to any of the polymorphs
with as-yet undetermined structures (III, IV, or V), and its
relationship (if any) to AMH-VI is not yet clear due to the lack of
suitable experimental data. The published neutron powder diffraction
data for AMH-III, IV, and V are not of high
quality;\cite{loveday2004ammonia,nelmes_airapt} in particular the
published powder pattern of AMH-IV consists only of very broad
reflections.
We have therefore carried out our own neutron powder diffraction
study,\cite{fortes2010report} with the aim of obtaining high
resolution data from these polymorphs. The results acquired so far
confirm that none of these polymorphs is likely to be the $P4/nmm$
phase.  There is, however, a discernible relationship between the
structures of the $P4/nmm$ phase and AMH-VI, as described below.  It
is worth noting that apparently small structural changes, particularly
those that break a crystal symmetry, can significantly affect a
diffraction pattern. Therefore it is reasonable to expect the
diffraction patterns of $P4/nmm$ and AMH-VI to differ, despite the
relationship between the structures.

\begin{figure}  
\includegraphics*[width=0.40\textwidth]{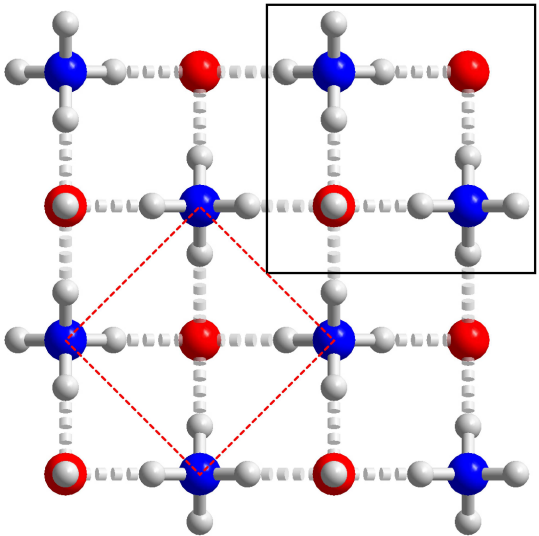}  
\caption{\label{fig:ionic_comp} The ionic $P4/nmm$
  phase viewed along the {\bf c}-axis.  This structure comprises
  sheets of ions hydrogen-bonded together (dashed rods) donated by the
  ammonium ions to the hydroxide.  The hydrogen atom of the hydroxide
  ion points directly along the {\bf c}-axis and does not appear to
  participate in a hydrogen bond.  Hence, the layers are not H-bonded
  to one another and consequently the structure exhibits a large
  compressibility along the {\bf c}-axis.  The dashed line shows the
  basis of a cubic structure that may be derived from the tetragonal
  cell (marked by the solid black line) by a small distortion, as
  discussed in the text.}
\end{figure}

With reference to Fig.\ \ref{fig:ionic_comp}, the hydrogen-bonded
layers in the $P4/nmm$ ionic phase form a network with a square motif,
which defines the tetragonal unit cell (marked in black). However, an
oblique cell (dashed line) is also marked that closely approximates a
cube having ammonium ions at the corners and a hydroxide ion near the
center. A slight shrinkage of the tetragonal {\bf a}- and {\bf b}-axes
($\sim$4.4\% relative to the value given in Table
\ref{table:structures}) while keeping the length of the {\bf c}-axis
unchanged, forms a perfect cube. Furthermore, shifting the fractional
$z$-coordinate of the oxygen atom from 0.6562 to 0.5 yields a
heavy-atom structure of space-group $Pm\bar{3}m$, {\bf a} = 3.3850
\AA, and fractional atomic coordinates N = 0, 0, 0, and O = 0.5, 0.5,
0.5.  Finally, mixing the occupancies of these two sites with ammonium
and hydroxyl ions gives an ionic equivalent of the AMH-VI structure.
In fact, the very small differences in computed Bragg intensities
between the ionic and molecular forms of AMH-VI lead us to conclude
that AMH-VI may well be ionic rather than molecular, and that the
$P4/nmm$ phase might simply be an ordered variant.

\begin{figure}
\includegraphics*[width=0.4\textwidth]{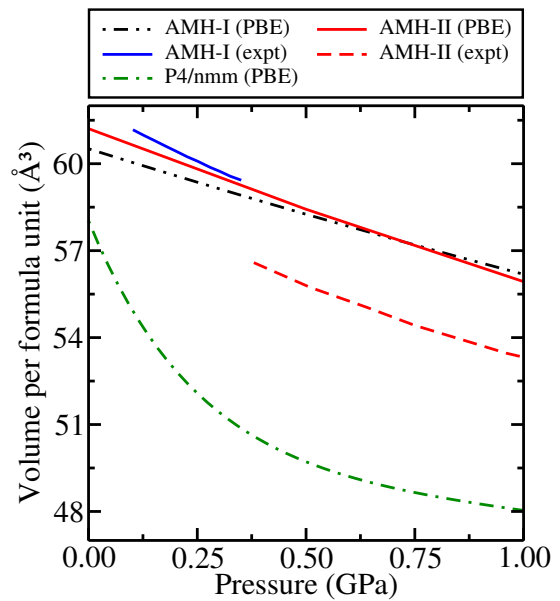}  
\caption{\label{fig:amh_vol} Volume-pressure relationships for the
  AMH-I and AMH-II phases and the new ionic phase of AMH reported
  here, calculated with the PBE functional, including the effects of
  ZP motion and temperature (175 K).  The experimental data were
  obtained from deuterated samples at 180 K at the Institut Laue
  Langevin, in the gas phase with a high-resolution powder
  diffractometer, and from a Paris-Edinburgh cell study with a
  high-intensity powder diffractometer, for AMH-I and AMH-II,
  respectively. }
\end{figure}

\section{Zero-point motion and finite temperature effects}
\label{sec:phonons}

The relatively small mass of hydrogen leads to large ZP motions in
AMH, where 5 out of every 7 nuclei are protons.  The ZP motion in AMH
may lead to important differences in the relative stabilities of the
phases, particularly when comparing a dense ionic phase with a less
dense molecular one.  We have investigated the effects of ZP motion on
the phase diagram of AMH within the quasi-harmonic approximation with
the supercell method and finite atomic displacements.  The
quasi-harmonic approximation normally gives a reasonable description
of vibrational effects, including thermal expansion.

We calculated the phonons of AMH-I and $P4/nmm$ in $112$ atom
supercells, while a set of finite displacement phonon calculations
were performed for AMH-II, which has a $112$ atom primitive cell.

Care was taken to ensure that the structures were very well relaxed
prior to performing phonon calculations, ensuring that any stresses on
the unit cell were less than $0.01$ eV/{\AA}$^{3}$ and that the forces
were converged to within $0.005$ eV/{\AA} ($0.003$ eV/{\AA} for
AMH-II). In addition, the fine grid on which the augmentation charge
density for the ultrasoft pseudopotentials is representated was
increased to $2.75$ times the multiple of the wavefunction grid to
obtain higher accuracy.

\begin{table*}  
\begin{ruledtabular}  
\begin{tabular}{lllll}  
  Structure & Method           & $K$ (GPa) & $K^{\prime}$ & $V_0$ (\AA$^3$)  \\\hline
  AMH-I ($Z=4$) & PBE              & \phantom{0}9.7 & \phantom{0}5.0  & 245.04 \\
            & PBE+ZP (0 K)     & 12.5 & \phantom{0}2.4 & 241.50 \\ 
            & PBE+ZP (175 K)   & 12.2 & \phantom{0}2.4 & 242.17 \\
            & PBE+G06          & 12.5 & \phantom{0}6.3 &  218.10\\
            & Experiment (140 K)\cite{loveday2004ammonia}&\phantom{0}8.9(4) & \phantom{0}4.2(3)  & 247.66  \\
            & Experiment (180 K)\cite{fortes2009crystal}&\phantom{0}7.33(3)&\phantom{0}5.3&248.00(2) \\
            & & & & \\
  AMH-II ($Z=16$)  & PBE              &\phantom{1}8.1 & \phantom{0}5.5 & 971.60  \\
            & PBE+ZP (0 K)     &\phantom{1}9.6 & \phantom{0}3.1 & 973.22 \\ 
            & PBE+ZP (175 K)   &\phantom{1}9.2 & \phantom{0}3.0 & 983.25 \\
            & PBE+G06          & 13.4 & \phantom{0}4.2 & 842.70 \\
            & Experiment (180 K)&\phantom{1}7.2(3) &\phantom{0}5.3(2) & 947(2) \\
            & & & & \\
  $P4/nmm$ ($Z=2$) & PBE              &\phantom{1}3.8 &13.8  & 106.51 \\
            & PBE+ZP (0 K)     &\phantom{1}4.6 &\phantom{1}7.4 & 113.13\\ 
            & PBE+ZP (175 K)   &\phantom{1}4.5 &\phantom{1}5.4  & 117.95 \\
            & PBE+G06          & 20.8          &11.0  &\phantom{1}83.01 \\
\end{tabular}  
\end{ruledtabular}
\caption{\label{table:properties} {The bulk modulus ($K$), the first 
    pressure derivative of the bulk modulus ($K^{\prime}$), and the 
    equilibrium volume ($V_0$). Fitting ranges: AMH-I 0--2.5 GPa, AMH-II 0.4--2.5 GPa. 
    $P4/nmm$  PBE 3--6 GPa, PBE+G06 0--2.5 GPa. } }
\end{table*}

\begin{table*}  
\begin{ruledtabular}  
\begin{tabular}{clllllllll}  
  Functional       & \multicolumn{3}{c}{Lattice parameters}          & &&Wyckoff&&& \\  
  & \multicolumn{3}{c}{(\AA, $^{\circ}$)}            & Atom& Site& symbol& \multicolumn{3}{l}{Fractional atomic coordinates}  \\\hline  
  PBE              & $a$=5.006     & $b$=5.006     & $c$=3.385       & N &$-42$ &2a&0.25 &0.7500 & 0.0000  \\  
  & $\alpha$=90    & $\beta$=90    & $\gamma$=90                     & O &$4mm$ &2c&0.25 &0.2500 & 0.6562  \\  
  &                &               &                                 & H1&$4mm$ &2c &0.75 &0.7500 & 0.6308  \\  
  &                &               &                                 & H2&$.m.$ &8i &0.25 &0.5727 & 0.8262  \\ 
\end{tabular}  
\end{ruledtabular}
\caption{\label{table:structures} {Structure of the $P4/nmm$ ionic 
    phase at $3$ GPa with the PBE functional (primitive cell, $Z=2$).} 
}
\end{table*}

\begin{table*}  
\begin{ruledtabular}  
\begin{tabular}{l|llll}  
Method             & \multicolumn{3}{c}{Transition Pressure (GPa)}          & \\
                   & AMH I $\rightarrow$ AMH II & AMH II $\rightarrow$ $P4/nmm$ & AMH I $\rightarrow$ $P4/nmm$ &  \\ \hline  
PBE                & \phantom{0}2.7                        & \phantom{0}2.8        & \phantom{0}2.8$^\dag$   &  \\  
PBE+G06            & \phantom{0}0.9$^\dag$                  & \phantom{0}0.4$^\dag$  & \phantom{0}0.5         &  \\ 
PBE+ZP (175 K)     & \phantom{0}2.7                        & \phantom{0}3.3        & \phantom{0}3.3$^\dag$   &  \\  
PBE0               & \phantom{0}n/a                        & \phantom{0}n/a        & 10.8                   &  \\ 
PBE0+ZP+G06 (0 K)  & \phantom{0}n/a                        & \phantom{0}n/a        & \phantom{0}8.8         &  \\ 
Experiment (180 K)\cite{fortes2009crystal} & \phantom{0}0.5                        & \phantom{0}n/a        & \phantom{0}n/a         &  \\
\end{tabular}  
\end{ruledtabular}
\caption{\label{table:transitions} {Summary of the transition pressures between 
    the AMH phases studied here, with each of the methods utilised. Transitions 
    which would occur at a pressure where neither phase is thermodynamically stable are denoted by 
    $^\dag$. Note that PBE0 calculations were not performed for the AMH II phase 
    and are therefore not included in the table (n/a).} 
}
\end{table*}

The Gibbs free energy at $175$ K is plotted against pressure for
AMH-I, AMH-II and the $P4/nmm$ ionic phase in Fig.\
\ref{fig:pbe_merged}, as calculated with the PBE functional. The
phonon pressure
was evaluated from the derivative of the ZP energy 
(or Helmholtz free energy at finite temperature) with respect to
volume.
The total pressure is the sum of the static DFT and phonon pressures.
The larger density of the ionic $P4/nmm$ phase leads to higher phonon
frequencies and hence destabilization relative to the molecular
phases.  The pressure obtained within PBE at which $P4/nmm$ becomes
the most stable at $175$ K is increased by $0.55$ GPa to $3.32$ GPa.
The inclusion of vibrational effects increases the pressure interval
of stability of AMH-II to about $0.7$ GPa, although this is still
smaller than the experimental interval of about 1.7
GPa.\cite{loveday2004ammonia,fortes2009crystal}

Experimentally, the transition from AMH-I to AMH-II at $\sim$0.35 GPa
results in a volume decrease of 4.6\%.\cite{10.1063/1.3245858} Both
PBE, and PBE with ZP motion at 175 K, give only a 2\% decrease in
volume at this transition, whereas PBE with the G06 dispersion
correction gives a decrease of 3.4\%.

Data from fits of the calculated pressure-volume data to the
third-order Birch-Murnaghan equation of state,\cite{birch1947finite}
with and without the ZP motion and temperature contributions, are
shown in Table \ref{table:properties}.  The uncorrected PBE results
compare most favorably with the experimental equilibrium volume, bulk
modulus and the first pressure derivative of the bulk modulus for both
AMH-I and AMH-II. It is interesting to note that there is one
exception to the expected increase in volume from including ZP motion,
AMH-I is seen to shrink slightly when ZP motion is taken into account.
All three structures are found to increase in volume when thermal
effects are included at 175 K, accompanied by a small reduction in the
bulk moduli relative to the values obtained on including ZP motion at
0 K.

\begin{figure}
\includegraphics*[width=0.4\textwidth]{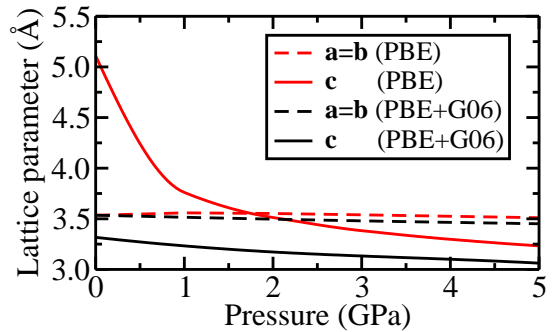}  
\caption{\label{fig:p4nmm_abc} Variation of the lattice
  parameters with pressure for the $P4/nmm$ ionic phase, calculated
  with the PBE functional, with and without the G06 dispersion
  correction.}
\end{figure}


\section{Dispersion correction}
\label{sec:g06}

The transition pressure between the AMH-I and II molecular phases
obtained with PBE of 2.8 GPa is significantly larger than the
experimental value of 0.5 GPa.  Here we explore the effects of
including dispersion forces which are not described by density
functionals such as PBE.  For this purpose we have recalculated the
phase diagram using the PBE functional with the Grimme semi-empirical
dispersion correction (G06),\cite{grimme2006semiempirical} see Fig.\
\ref{fig:pbe_merged} b).
The significant overestimate of the transition pressure obtained with
PBE is substantially improved by including the G06 correction.  The
transition pressure between the molecular AMH-I and AMH-II phases is
reduced to $1$ GPa using the PBE+G06 functional, which is still
somewhat larger than the experimental value of $0.5$ GPa.  However,
the G06 correction significantly favours the denser $P4/nmm$ ionic
phase, which now becomes stable at about $0.5$ GPa. Each phase
undergoes a substantial volume contraction when the G06 correction is
included, which is likely due to an overestimation of dispersion
effects.

The large effect of the dispersion correction on the ionic structure
may seem paradoxical as it amounts to a relatively small fraction of
the binding energy of the NH$^+_4\cdots$OH$^-$ ionic complex (see
Fig.\ \ref{fig:Eint-all-ion-1img}).  The rapid reduction in the volume
of $P4/nmm$ with applied pressure apparent in Fig.\ \ref{fig:amh_vol}
is almost entirely associated with a contraction along the {\bf c}
lattice parameter.  As shown in Fig.\ \ref{fig:ionic_comp}, the
hydrogen bonds form a square net within the {\bf a}-{\bf b}-planes of
$P4/nmm$, but there is no apparent hydrogen bonding between the layers
(see Fig.\ \ref{fig:ionic_struct}), and therefore it is very soft in
the {\bf c} direction.

  


\section{Beyond the PBE functional}
\label{sec:functional}

\subsection{Calculations for molecular and ionic fragments}

Despite giving a good description of the bulk structural properties of
AMH-I and II, the PBE functional leads to substantially incorrect
transition pressures.  PBE overestimates the pressure of the AMH-I/II
transition, and the $P4/nmm$ ionic phase is predicted to be
sufficiently stable to virtually eliminate the region of stability of
AMH-II.  Even accounting for ZP motion and thermal effects it
is hard to reconcile the apparent stability of the $P4/nmm$ ionic
phase with the existing room temperature experiments in which the
ionic AMH-VI phase was not found at pressures as high as $9$
GPa.\cite{fortes2011iucr} The structures of AMH-III, IV and V have not
yet been solved, and one or more of these phases may be ionic.

\begin{figure}  
\includegraphics*[width=0.45\textwidth]{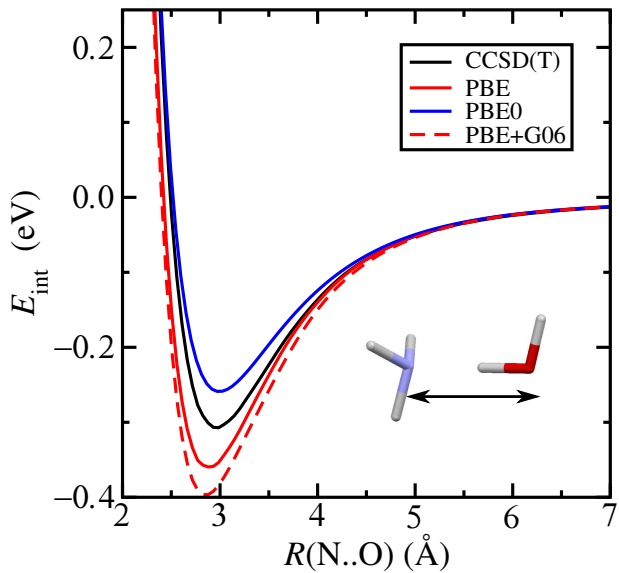}  
\caption{\label{fig:Eint-all-mol_1img} Interaction
  energy against separation of NH$_3$ and H$_2$O for a configuration
  taken from the AMH-I molecular phase. }
\end{figure}

\begin{figure}  
\includegraphics*[width=0.45\textwidth]{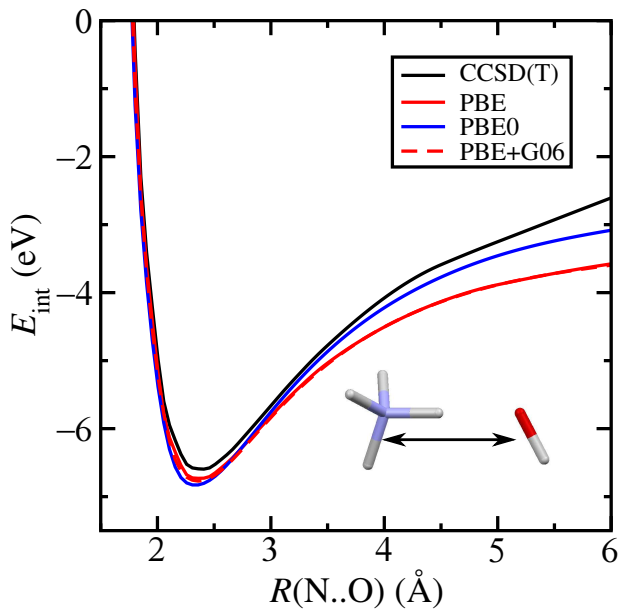}  
\caption{\label{fig:Eint-all-ion-1img} Interaction
  energy against separation of NH$^+_4$ and OH$^-$ for a configuration
  taken from the $P4/nmm$ ionic phase. }
\end{figure}

To obtain insight into this problem we have selected fragments of the
AMH-I and $P4/nmm$ crystals (at pressures of $1$ and $4$ GPa,
respectively) for a more detailed analysis.  These fragments were
chosen to be representative of the interactions in the crystal and
consisted of pairs of molecules/ions in close proximity: NH$_3$ and
H$_2$O for the AHM-I crystal (Fig.\ \ref{fig:Eint-all-mol_1img}) and
NH$_4^+$ and OH$^-$ ions for the $P4/nmm$ crystal (Fig.\
\ref{fig:Eint-all-ion-1img}).  In Figs.\ \ref{fig:Eint-all-mol_1img}
and \ref{fig:Eint-all-ion-1img} we plot interaction energies for these
pairs calculated using the supermolecular approach, that is, $E_{\rm
  int} = E(AB) - E(A) - E(B)$, where $E(A)$ and $E(B)$ are the
energies of the monomers and $E(AB)$ is the energy of the complex. We
have used the CCSD(T), MP2, PBE, PBE+G06 and PBE0 methods, though the
MP2 results are not shown as they are essentially identical to the
CCSD(T) values. We have calculated counterpoise-corrected interaction
energies using the Boys and Bernardi\cite{BoysB70} scheme with a
Sadlej-pVTZ basis\cite{Sadlej91} augmented with a small set of ``bond
centered functions''\cite{BukowskiSJJSKWR99} which help to saturate
the dispersion energy.\cite{WilliamsMSJ95} Selected calculations with
the larger aug-cc-pVTZ basis set suggest that the CCSD(T) interaction
energies are converged to 5\% at the equilibrium geometry and better
at larger separations.  These calculations were performed using the
{\sc DALTON 2.0} program.\cite{DALTON2} For deeper insight into the
nature of the interaction energies we have additionally used the {\sc
  CamCASP} program\cite{CamCASP} to perform symmetry adapted
perturbation theory (SAPT)
DFT\cite{MisquittaS02,MisquittaJS03,MisquittaPJS05b} calculations to
decompose the interaction energies into physical components.

Fig.\ \ref{fig:Eint-all-mol_1img} shows that PBE overbinds the
NH$_3$$\cdots$H$_2$O complex, although the equilibrium bond length
agrees reasonably well with the CCSD(T) value. The overestimation of
the bulk modulus ($K$) of AMH-I reported in Table
\ref{table:properties} may be a reflection of the overbinding of PBE
and the consequent overestimate of the curvature of the potential
well. In contrast, PBE0 underbinds the complex, but results in an
equilibrium separation in near perfect agreement with the CCSD(T)
calculations. The SAPT(DFT) energy decomposition shows that
asymptotically the NH$_3$$\cdots$H$_2$O interaction is dominated by
the dipole-dipole electrostatic energies with the dispersion and
polarization energies being negligible. Consequently it should not be
a surprise that both the correlated and density-functional methods
agree in this region.

The ionic system behaves very differently from the molecular one. Note
that the interaction energies for the ionic system shown in Fig.\
\ref{fig:Eint-all-ion-1img} are an order of magnitude larger than in
the molecular system. The SAPT(DFT) results show that the energetics
are almost completely dominated by the charge-charge electrostatic
interaction between the ions, with the dispersion energy contribution
being insignificant even at the relatively short N$\cdots$O separation
of 2.5 \AA.  This is exactly the kind of system for which local and
semi-local density functionals are expected to achieve high accuracy.
Indeed, the PBE and PBE0 energies are close to the CCSD(T) result at
the equilibrium geometry and on the repulsive wall.  In contrast to
the molecular system, however, PBE significantly overbinds the ionic
complex at large separations. As can be seen in Fig.\
\ref{fig:Eint-all-ion-1img}, the overbinding compared with CCSD(T) is
20\% at 5 \AA\ but it grows to more than 50\% at 6.5 \AA.  The
interaction energies are still relatively large at these separations,
and consequently these errors have important effects within the
crystal. This is consistent with our observation that the $P4/nmm$
ionic phase is over-stabilized with respect to the molecular phases.

A partial charge analysis shows that the substantial difference
between the PBE and CCSD(T) interaction energies at large separations
is a consequence of excessive charge transfer.  Within PBE the
magnitude of the charges on the monomers grows as they are pulled
apart, becoming as large as $\pm1.2e$ at a N$\cdots$O separation of
6.5 \AA.  In contrast, the CCSD(T) potential energy curve can be
fitted very well with a $-1/R$ form corresponding to partial charges
of $\pm1e$.  The PBE charge transfer error is analogous to the
delocalization or static correlation errors exhibited by local and
semi-local density functionals.\cite{CohenM-SY08a} This error can be
at least partially corrected by introducing some fraction of non-local
exchange.  Using PBE0 gives a significant improvement upon the PBE
energies, with the overbinding compared with CCSD(T) at 5 \AA\ and 6.5
\AA\ reduced to 6\% and 28\%, respectively, see Fig.\
\ref{fig:Eint-all-ion-1img}.

\subsection{PBE0 calculations for crystalline AMH}


As the PBE0 functional appears to offer a partial solution to the
charge transfer errors which arise with PBE, we performed calculations
for the molecular AMH-I and ionic $P4/nmm$ phases using the PBE0
hybrid functional.
Because of the slow convergence of the exchange terms with distance,
the PBE0 calculations for the crystal were as much as two orders of
magnitude more computationally expensive than the corresponding PBE
calculations. For this reason geometry optimizations were not possible
with PBE0 and instead we performed single-point energy calculations
using the relaxed PBE structures.  We attempted to obtain equation of
state parameters from the PBE0 results for AMH-I and $P4/nmm$.  The
calculated energies were fitted to the third-order Birch-Murnaghan
equation of state and the pressure was obtained by differentiation. We
obtained equilibrium volumes of $V_0=220$ \AA$^3$ for AMH-I and
$V_0=172$ \AA$^3$ for $P4/nmm$, which are similar to those obtained
from the PBE+G06 calculations.  We were not able to obtain reliable
estimates of $K$ or $K^{\prime}$ from our PBE0 data as the values were
sensitive to the data and fitting procedure.  The use of the PBE
structures for the PBE0 calculations is an approximation.  However, we
believe this approach to be reasonably robust because enthalpies
obtained using partially relaxed PBE structures were almost identical
to those from the fully relaxed structures.


Mulliken charge analysis of the $P4/nmm$ crystal suggests that the
electronic charge transfer between the ``OH$^-$'' and ``NH$_4^+$''
ions has a magnitude of about $0.70e$ when using the PBE functional
and $0.55e$ with PBE0.  The improved description of the ionic phase
provided by PBE0 destabilizes the $P4/nmm$ ionic phase relative to
AMH-I, and consequently $P4/nmm$ enters above $10.8$ GPa, as can be
seen in Fig.\ \ref{fig:pbe0_merged}. AMH-II, AMH-IV and AMH-VI are not
present on this phase diagram, but they are expected to have regions
of stability between AMH-I and $P4/nmm$, pushing the entry of the new
ionic phase to even higher pressures. The inclusion of vibrational
effects further destabilizes $P4/nmm$ and it only becomes stable at
$11.3$ GPa, although conversely the inclusion of the G06 dispersion
correction reduces the transition pressure to $8.8$ GPa, as shown in
Fig.\ \ref{fig:pbe0_merged}.

\section{Band gaps of the phases}

\begin{figure}  
\includegraphics*[width=0.4\textwidth]{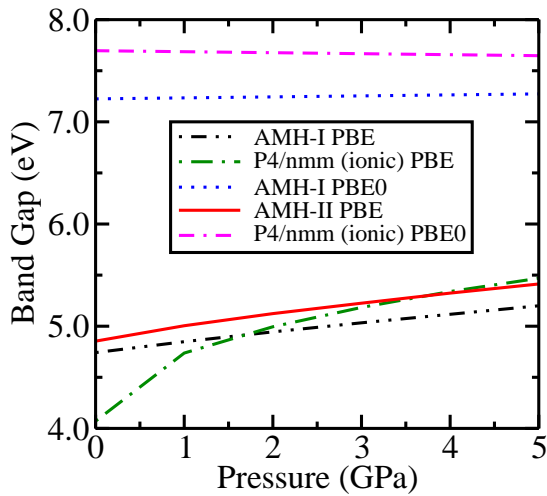}  
\caption{\label{fig:bg_pbe} Variation of the band gaps
  of the phases with pressure. AMH-I (and AMH-II using PBE only) have
  direct band gaps, while both the PBE and PBE0 calculations predict
  that the $P4/nmm$ phase has an indirect band gap. The PBE0
  calculations were performed using the PBE structures.}
\end{figure}

The minimum band gaps obtained with the PBE and PBE0 functionals are
shown in Fig.\ \ref{fig:bg_pbe} calculated using the {\sc LinDOS}
code.\cite{lindos} While the molecular structures have direct band
gaps, both PBE and PBE0 calculations predict $P4/nmm$ to have an
indirect band gap. The band gaps provided by the PBE0 functional,
which includes explicit exchange interactions, are substantially
larger than those predicted by PBE.

\section{Conclusions}


We have performed extensive \emph{ab initio} searches for new phases
of AMH at pressures up to 12 GPa using the PBE functional. A new ionic
structure of space group $P4/nmm$ was found to be stable above 2.8
GPa. Subsequent investigations into the effects of temperature, ZP
motion and dispersion forces found that the latter two play a
substantial role in determining the relative stabilities of the
phases. The inclusion of ZP motion destabilizes the dense $P4/nmm$
ionic phase; conversely the dispersion correction leads to a
significant underestimation of the transition pressure from molecular
AMH to ionic $P4/nmm$. The dispersion forces are attractive and
therefore tend to favour denser structures.

The relationship described earlier between the structures of the
$P4/nmm$ phase and AMH-VI may indicate that we have discovered an
ordered ionic variant of AMH-VI.  Alternatively, the $P4/nmm$ phase
may be related to one of the three experimentally observed AMH phases
(III, IV, V) whose structures are unknown.  As the $P4/nmm$ phase was
the only new structure found which is predicted to be
thermodynamically stable within some pressure range, it is likely that
if it is one of the three unknown AMH structures, then the other two
will have unit cells with $Z>8$.

Accurate CCSD(T) calculations on representative complexes from the
ionic and molecular AMH phases revealed that the PBE functional
substantially overbinds the ionic phase. We have presented evidence
that this overbinding arises from an overestimate of the electronic
charge transfer accompanying the proton transfer, which can be
partially remedied by using the hybrid PBE0 functional.  Using the
PBE0 functional leads to an increase in the enthalpy of the ionic
$P4/nmm$ phase by about 0.6 eV per f.u.\ relative to AMH-I.  The
transition pressure from AMH-I to $P4/nmm$ phase is substantially
increased, which eliminates the inconsistency with experiment.  This
failure of the PBE functional, for an electrostatic-bound system for
which GGA-type density functionals are typically assumed to be
accurate, is likely to have implications for other systems.  Further
experimental work is necessary to explore the phase diagram of AMH at
9 GPa and above to confirm our assignment of $P4/nmm$ as a stable
ionic phase of AMH.


\section{Acknowledgments}

This work was supported by the Engineering and Physical Research
Council (EPSRC) of the UK. Computational resources were provided by
the Cambridge High Performance Computing Service.  ADF acknowledges
funding from the Science and Technology Facilities Council (STFC), UK,
fellowship number PP/E006515/1.


%





\begin{thebibliography}{42}%
\makeatletter
\providecommand \@ifxundefined [1]{%
 \@ifx{#1\undefined}
}%
\providecommand \@ifnum [1]{%
 \ifnum #1\expandafter \@firstoftwo
 \else \expandafter \@secondoftwo
 \fi
}%
\providecommand \@ifx [1]{%
 \ifx #1\expandafter \@firstoftwo
 \else \expandafter \@secondoftwo
 \fi
}%
\providecommand \natexlab [1]{#1}%
\providecommand \enquote  [1]{``#1''}%
\providecommand \bibnamefont  [1]{#1}%
\providecommand \bibfnamefont [1]{#1}%
\providecommand \citenamefont [1]{#1}%
\providecommand \href@noop [0]{\@secondoftwo}%
\providecommand \href [0]{\begingroup \@sanitize@url \@href}%
\providecommand \@href[1]{\@@startlink{#1}\@@href}%
\providecommand \@@href[1]{\endgroup#1\@@endlink}%
\providecommand \@sanitize@url [0]{\catcode `\\12\catcode `\$12\catcode
  `\&12\catcode `\#12\catcode `\^12\catcode `\_12\catcode `\%12\relax}%
\providecommand \@@startlink[1]{}%
\providecommand \@@endlink[0]{}%
\providecommand \url  [0]{\begingroup\@sanitize@url \@url }%
\providecommand \@url [1]{\endgroup\@href {#1}{\urlprefix }}%
\providecommand \urlprefix  [0]{URL }%
\providecommand \Eprint [0]{\href }%
\@ifxundefined \urlstyle {%
  \providecommand \doi  [0]{\begingroup \@sanitize@url \@doi}%
  \providecommand \@doi [1]{\endgroup \@@startlink {\doibase
  #1}doi:\discretionary {}{}{}#1\@@endlink }%
}{%
  \providecommand \doi  [0]{doi:\discretionary{}{}{}\begingroup
  \urlstyle{rm}\Url }%
}%
\providecommand \doibase [0]{http://dx.doi.org/}%
\providecommand \Doi [0]{\begingroup \@sanitize@url \@Doi }%
\providecommand \@Doi  [1]{\endgroup\@@startlink{\doibase#1}\@@Doi}%
\providecommand \@@Doi [1]{#1\@@endlink}%
\providecommand \selectlanguage [0]{\@gobble}%
\providecommand \bibinfo  [0]{\@secondoftwo}%
\providecommand \bibfield  [0]{\@secondoftwo}%
\providecommand \translation [1]{[#1]}%
\providecommand \BibitemOpen [0]{}%
\providecommand \bibitemStop [0]{}%
\providecommand \bibitemNoStop [0]{.\EOS\space}%
\providecommand \EOS [0]{\spacefactor3000\relax}%
\providecommand \BibitemShut  [1]{\csname bibitem#1\endcsname}%
\bibitem [{\citenamefont {Loveday}\ and\ \citenamefont
  {Nelmes}(2004)}]{loveday2004ammonia}%
  \BibitemOpen
  \bibfield  {author} {\bibinfo {author} {\bibfnamefont {J.~S.}\ \bibnamefont
  {Loveday}}\ and\ \bibinfo {author} {\bibfnamefont {R.~J.}\ \bibnamefont
  {Nelmes}},\ }\href
  {http://www.tandfonline.com/doi/abs/10.1080/08957950410001661990} {\bibfield
  {journal} {\bibinfo  {journal} {High Press. Res.},\ }\textbf {\bibinfo
  {volume} {24}},\ \bibinfo {pages} {45} (\bibinfo {year} {2004})}\BibitemShut
  {NoStop}%
\bibitem [{\citenamefont {Olovsson}\ and\ \citenamefont
  {Templeton}(1959)}]{olovsson1959crystal}%
  \BibitemOpen
  \bibfield  {author} {\bibinfo {author} {\bibfnamefont {I.}~\bibnamefont
  {Olovsson}}\ and\ \bibinfo {author} {\bibfnamefont {D.}~\bibnamefont
  {Templeton}},\ }\href
  {http://scripts.iucr.org/cgi-bin/paper?S0365110X59002419} {\bibfield
  {journal} {\bibinfo  {journal} {Acta Cryst.},\ }\textbf {\bibinfo {volume}
  {12}},\ \bibinfo {pages} {827} (\bibinfo {year} {1959})}\BibitemShut
  {NoStop}%
\bibitem [{\citenamefont {Loveday}\ and\ \citenamefont
  {Nelmes}(2000)}]{nelmes_airapt}%
  \BibitemOpen
  \bibfield  {author} {\bibinfo {author} {\bibfnamefont {J.~S.}\ \bibnamefont
  {Loveday}}\ and\ \bibinfo {author} {\bibfnamefont {R.~J.}\ \bibnamefont
  {Nelmes}},\ }\href@noop {} {\bibfield  {journal} {\bibinfo  {journal} {Sci.
  Tech. High Press. Proc. AIRAPT-17},\ \bibinfo {pages} {133}} (\bibinfo {year}
  {2000})}\BibitemShut {NoStop}%
\bibitem [{\citenamefont {Loveday}\ and\ \citenamefont
  {Nelmes}(1999)}]{loveday1999ammonia}%
  \BibitemOpen
  \bibfield  {author} {\bibinfo {author} {\bibfnamefont {J.~S.}\ \bibnamefont
  {Loveday}}\ and\ \bibinfo {author} {\bibfnamefont {R.~J.}\ \bibnamefont
  {Nelmes}},\ }\href {http://prl.aps.org/abstract/PRL/v83/i21/p4329_1}
  {\bibfield  {journal} {\bibinfo  {journal} {Phys. Rev. Lett.},\ }\textbf
  {\bibinfo {volume} {83}},\ \bibinfo {pages} {4329} (\bibinfo {year}
  {1999})}\BibitemShut {NoStop}%
\bibitem [{\citenamefont {Pickard}\ and\ \citenamefont
  {Needs}(2011)}]{airss_review_2011}%
  \BibitemOpen
  \bibfield  {author} {\bibinfo {author} {\bibfnamefont {C.~J.}\ \bibnamefont
  {Pickard}}\ and\ \bibinfo {author} {\bibfnamefont {R.~J.}\ \bibnamefont
  {Needs}},\ }\href {http://stacks.iop.org/0953-8984/23/i=5/a=053201}
  {\bibfield  {journal} {\bibinfo  {journal} {J. Phys.: Condens. Matter},\
  }\textbf {\bibinfo {volume} {23}},\ \bibinfo {pages} {053201} (\bibinfo
  {year} {2011})}\BibitemShut {NoStop}%
\bibitem [{\citenamefont {Pickard}\ and\ \citenamefont
  {Needs}(2006)}]{pickard:045504}%
  \BibitemOpen
  \bibfield  {author} {\bibinfo {author} {\bibfnamefont {C.~J.}\ \bibnamefont
  {Pickard}}\ and\ \bibinfo {author} {\bibfnamefont {R.~J.}\ \bibnamefont
  {Needs}},\ }\Doi {10.1103/PhysRevLett.97.045504} {\bibfield  {journal}
  {\bibinfo  {journal} {Phys. Rev. Lett.},\ }\textbf {\bibinfo {volume} {97}},\
  \bibinfo {eid} {045504} (\bibinfo {year} {2006})}\BibitemShut {NoStop}%
\bibitem [{\citenamefont {Fortes}\ \emph
  {et~al.}(2009){\natexlab{a}}\citenamefont {Fortes}, \citenamefont {Suard},
  \citenamefont {Lem\'{e}e-Cailleau}, \citenamefont {Pickard},\ and\
  \citenamefont {Needs}}]{fortes2009crystal}%
  \BibitemOpen
  \bibfield  {author} {\bibinfo {author} {\bibfnamefont {A.~D.}\ \bibnamefont
  {Fortes}}, \bibinfo {author} {\bibfnamefont {E.}~\bibnamefont {Suard}},
  \bibinfo {author} {\bibfnamefont {M.~H.}\ \bibnamefont {Lem\'{e}e-Cailleau}},
  \bibinfo {author} {\bibfnamefont {C.~J.}\ \bibnamefont {Pickard}}, \ and\
  \bibinfo {author} {\bibfnamefont {R.~J.}\ \bibnamefont {Needs}},\ }\href
  {http://pubs.acs.org/doi/abs/10.1021/ja9052569} {\bibfield  {journal}
  {\bibinfo  {journal} {J. Am. Chem. Soc.},\ }\textbf {\bibinfo {volume}
  {131}},\ \bibinfo {pages} {13508} (\bibinfo {year}
  {2009}{\natexlab{a}})}\BibitemShut {NoStop}%
\bibitem [{\citenamefont {Fortes}\ \emph
  {et~al.}(2009){\natexlab{b}}\citenamefont {Fortes}, \citenamefont {Suard},
  \citenamefont {Lem\'{e}e-Cailleau}, \citenamefont {Pickard},\ and\
  \citenamefont {Needs}}]{10.1063/1.3245858}%
  \BibitemOpen
  \bibfield  {author} {\bibinfo {author} {\bibfnamefont {A.~D.}\ \bibnamefont
  {Fortes}}, \bibinfo {author} {\bibfnamefont {E.}~\bibnamefont {Suard}},
  \bibinfo {author} {\bibfnamefont {M.~H.}\ \bibnamefont {Lem\'{e}e-Cailleau}},
  \bibinfo {author} {\bibfnamefont {C.~J.}\ \bibnamefont {Pickard}}, \ and\
  \bibinfo {author} {\bibfnamefont {R.~J.}\ \bibnamefont {Needs}},\ }\Doi
  {DOI:10.1063/1.3245858} {\bibfield  {journal} {\bibinfo  {journal} {J. Chem.
  Phys.},\ }\textbf {\bibinfo {volume} {131}},\ \bibinfo {pages} {154503}
  (\bibinfo {year} {2009}{\natexlab{b}})}\BibitemShut {NoStop}%
\bibitem [{\citenamefont {Kargel}(1992)}]{kargel1992ammonia}%
  \BibitemOpen
  \bibfield  {author} {\bibinfo {author} {\bibfnamefont {J.}~\bibnamefont
  {Kargel}},\ }\href
  {http://www.sciencedirect.com/science/article/pii/001910359290118Q}
  {\bibfield  {journal} {\bibinfo  {journal} {Icarus},\ }\textbf {\bibinfo
  {volume} {100}},\ \bibinfo {pages} {556} (\bibinfo {year}
  {1992})}\BibitemShut {NoStop}%
\bibitem [{\citenamefont {Fortes}\ and\ \citenamefont
  {Choukroun}(2010)}]{springerlink:10.1007/s11214-010-9633-3}%
  \BibitemOpen
  \bibfield  {author} {\bibinfo {author} {\bibfnamefont {A.~D.}\ \bibnamefont
  {Fortes}}\ and\ \bibinfo {author} {\bibfnamefont {M.}~\bibnamefont
  {Choukroun}},\ }\href {http://www.springerlink.com/content/w45mq8013g523276/}
  {\bibfield  {journal} {\bibinfo  {journal} {Space Sci. Rev.},\ }\textbf
  {\bibinfo {volume} {153}},\ \bibinfo {pages} {185} (\bibinfo {year}
  {2010})}\BibitemShut {NoStop}%
\bibitem [{\citenamefont {Fortes}\ \emph
  {et~al.}(2009){\natexlab{c}}\citenamefont {Fortes}, \citenamefont {Wood},
  \citenamefont {Vo\u{c}adlo}, \citenamefont {Knight}, \citenamefont
  {Marshall}, \citenamefont {Tucker},\ and\ \citenamefont
  {Fernandez-Alonso}}]{fortes2009phase}%
  \BibitemOpen
  \bibfield  {author} {\bibinfo {author} {\bibfnamefont {A.~D.}\ \bibnamefont
  {Fortes}}, \bibinfo {author} {\bibfnamefont {I.~G.}\ \bibnamefont {Wood}},
  \bibinfo {author} {\bibfnamefont {L.}~\bibnamefont {Vo\u{c}adlo}}, \bibinfo
  {author} {\bibfnamefont {K.~S.}\ \bibnamefont {Knight}}, \bibinfo {author}
  {\bibfnamefont {W.~G.}\ \bibnamefont {Marshall}}, \bibinfo {author}
  {\bibfnamefont {M.~G.}\ \bibnamefont {Tucker}}, \ and\ \bibinfo {author}
  {\bibfnamefont {F.}~\bibnamefont {Fernandez-Alonso}},\ }\href
  {http://scripts.iucr.org/cgi-bin/paper?ks5224} {\bibfield  {journal}
  {\bibinfo  {journal} {J. Appl. Cryst.},\ }\textbf {\bibinfo {volume} {42}},\
  \bibinfo {pages} {846} (\bibinfo {year} {2009}{\natexlab{c}})}\BibitemShut
  {NoStop}%
\bibitem [{\citenamefont {Fortes}(2011)}]{Fortes2011}%
  \BibitemOpen
  \bibfield  {author} {\bibinfo {author} {\bibfnamefont {A.~D.}\ \bibnamefont
  {Fortes}},\ }\href
  {http://www.sciencedirect.com/science/article/pii/S0032063311001401}
  {\bibfield  {journal} {\bibinfo  {journal} {Planet. Space Sci.},\ }\textbf
  {\bibinfo {volume} {In Press, Corrected Proof}},\  (\bibinfo {year}
  {2011})}\BibitemShut {NoStop}%
\bibitem [{\citenamefont {Lunine}\ and\ \citenamefont
  {Stevenson}(1987)}]{lunine1987clathrate}%
  \BibitemOpen
  \bibfield  {author} {\bibinfo {author} {\bibfnamefont {J.}~\bibnamefont
  {Lunine}}\ and\ \bibinfo {author} {\bibfnamefont {D.}~\bibnamefont
  {Stevenson}},\ }\href
  {http://www.sciencedirect.com/science/article/pii/0019103587900753}
  {\bibfield  {journal} {\bibinfo  {journal} {Icarus},\ }\textbf {\bibinfo
  {volume} {70}},\ \bibinfo {pages} {61} (\bibinfo {year} {1987})}\BibitemShut
  {NoStop}%
\bibitem [{\citenamefont {Fu}\ \emph {et~al.}(2010)\citenamefont {Fu},
  \citenamefont {O'Connell},\ and\ \citenamefont {Sasselov}}]{fu2010interior}%
  \BibitemOpen
  \bibfield  {author} {\bibinfo {author} {\bibfnamefont {R.}~\bibnamefont
  {Fu}}, \bibinfo {author} {\bibfnamefont {R.}~\bibnamefont {O'Connell}}, \
  and\ \bibinfo {author} {\bibfnamefont {D.}~\bibnamefont {Sasselov}},\ }\href
  {http://iopscience.iop.org/0004-637X/708/2/1326} {\bibfield  {journal}
  {\bibinfo  {journal} {ApJ},\ }\textbf {\bibinfo {volume} {708}},\ \bibinfo
  {pages} {1326} (\bibinfo {year} {2010})}\BibitemShut {NoStop}%
\bibitem [{\citenamefont {Hubbard}\ and\ \citenamefont
  {MacFarlane}(1980)}]{hubbard1980structure}%
  \BibitemOpen
  \bibfield  {author} {\bibinfo {author} {\bibfnamefont {W.}~\bibnamefont
  {Hubbard}}\ and\ \bibinfo {author} {\bibfnamefont {J.}~\bibnamefont
  {MacFarlane}},\ }\href
  {http://www.agu.org/pubs/crossref/1980/JB085iB01p00225.shtml} {\bibfield
  {journal} {\bibinfo  {journal} {J. Geophys. Res.},\ }\textbf {\bibinfo
  {volume} {85}},\ \bibinfo {pages} {225} (\bibinfo {year} {1980})}\BibitemShut
  {NoStop}%
\bibitem [{\citenamefont {Cavazzoni}\ \emph {et~al.}(1999)\citenamefont
  {Cavazzoni}, \citenamefont {Chiarotti}, \citenamefont {Scandolo},
  \citenamefont {Tosatti}, \citenamefont {Bernasconi},\ and\ \citenamefont
  {Parrinello}}]{cavazzoni1999superionic}%
  \BibitemOpen
  \bibfield  {author} {\bibinfo {author} {\bibfnamefont {C.}~\bibnamefont
  {Cavazzoni}}, \bibinfo {author} {\bibfnamefont {G.}~\bibnamefont
  {Chiarotti}}, \bibinfo {author} {\bibfnamefont {S.}~\bibnamefont {Scandolo}},
  \bibinfo {author} {\bibfnamefont {E.}~\bibnamefont {Tosatti}}, \bibinfo
  {author} {\bibfnamefont {M.}~\bibnamefont {Bernasconi}}, \ and\ \bibinfo
  {author} {\bibfnamefont {M.}~\bibnamefont {Parrinello}},\ }\href
  {http://www.sciencemag.org/content/283/5398/44.full} {\bibfield  {journal}
  {\bibinfo  {journal} {Science},\ }\textbf {\bibinfo {volume} {283}},\
  \bibinfo {pages} {44} (\bibinfo {year} {1999})}\BibitemShut {NoStop}%
\bibitem [{\citenamefont {Fortes}\ \emph {et~al.}(2001)\citenamefont {Fortes},
  \citenamefont {Brodholt}, \citenamefont {Wood}, \citenamefont {Vo\u{c}adlo},\
  and\ \citenamefont {Jenkins}}]{fortes2001ab}%
  \BibitemOpen
  \bibfield  {author} {\bibinfo {author} {\bibfnamefont {A.~D.}\ \bibnamefont
  {Fortes}}, \bibinfo {author} {\bibfnamefont {J.~P.}\ \bibnamefont
  {Brodholt}}, \bibinfo {author} {\bibfnamefont {I.~G.}\ \bibnamefont {Wood}},
  \bibinfo {author} {\bibfnamefont {L.}~\bibnamefont {Vo\u{c}adlo}}, \ and\
  \bibinfo {author} {\bibfnamefont {H.~D.~B.}\ \bibnamefont {Jenkins}},\ }\href
  {http://jcp.aip.org/resource/1/jcpsa6/v115/i15/p7006_s1} {\bibfield
  {journal} {\bibinfo  {journal} {J. Chem. Phys.},\ }\textbf {\bibinfo {volume}
  {115}},\ \bibinfo {pages} {7006} (\bibinfo {year} {2001})}\BibitemShut
  {NoStop}%
\bibitem [{\citenamefont {Fortes}(2004)}]{fortesphd}%
  \BibitemOpen
  \bibfield  {author} {\bibinfo {author} {\bibfnamefont {A.~D.}\ \bibnamefont
  {Fortes}},\ }\href
  {http://www.homepages.ucl.ac.uk/~ucfbanf/publications/thesis/thesis.htm}
  {Ph.D. thesis},\ \bibinfo  {school} {University of London} (\bibinfo {year}
  {2004})\BibitemShut {NoStop}%
\bibitem [{\citenamefont {Pickard}\ and\ \citenamefont
  {Needs}(2008)}]{pickard2008hca}%
  \BibitemOpen
  \bibfield  {author} {\bibinfo {author} {\bibfnamefont {C.~J.}\ \bibnamefont
  {Pickard}}\ and\ \bibinfo {author} {\bibfnamefont {R.~J.}\ \bibnamefont
  {Needs}},\ }\href
  {http://www.nature.com/nmat/journal/v7/n10/abs/nmat2261.html} {\bibfield
  {journal} {\bibinfo  {journal} {Nature Mater.},\ }\textbf {\bibinfo {volume}
  {7}},\ \bibinfo {pages} {775} (\bibinfo {year} {2008})}\BibitemShut {NoStop}%
\bibitem [{\citenamefont {Clark}\ \emph {et~al.}(2005)\citenamefont {Clark},
  \citenamefont {Segall}, \citenamefont {Pickard}, \citenamefont {Hasnip},
  \citenamefont {Probert}, \citenamefont {Refson},\ and\ \citenamefont
  {Payne}}]{clark2005first}%
  \BibitemOpen
  \bibfield  {author} {\bibinfo {author} {\bibfnamefont {S.~J.}\ \bibnamefont
  {Clark}}, \bibinfo {author} {\bibfnamefont {M.~D.}\ \bibnamefont {Segall}},
  \bibinfo {author} {\bibfnamefont {C.~J.}\ \bibnamefont {Pickard}}, \bibinfo
  {author} {\bibfnamefont {P.~J.}\ \bibnamefont {Hasnip}}, \bibinfo {author}
  {\bibfnamefont {M.~I.~J.}\ \bibnamefont {Probert}}, \bibinfo {author}
  {\bibfnamefont {K.}~\bibnamefont {Refson}}, \ and\ \bibinfo {author}
  {\bibfnamefont {M.~C.}\ \bibnamefont {Payne}},\ }\href@noop {} {\bibfield
  {journal} {\bibinfo  {journal} {Z. Kristallogr.},\ }\textbf {\bibinfo
  {volume} {220}},\ \bibinfo {pages} {567} (\bibinfo {year}
  {2005})}\BibitemShut {NoStop}%
\bibitem [{\citenamefont {Perdew}\ \emph {et~al.}(1996)\citenamefont {Perdew},
  \citenamefont {Burke},\ and\ \citenamefont
  {Ernzerhof}}]{PhysRevLett.77.3865}%
  \BibitemOpen
  \bibfield  {author} {\bibinfo {author} {\bibfnamefont {J.~P.}\ \bibnamefont
  {Perdew}}, \bibinfo {author} {\bibfnamefont {K.}~\bibnamefont {Burke}}, \
  and\ \bibinfo {author} {\bibfnamefont {M.}~\bibnamefont {Ernzerhof}},\ }\Doi
  {10.1103/PhysRevLett.77.3865} {\bibfield  {journal} {\bibinfo  {journal}
  {Phys. Rev. Lett.},\ }\textbf {\bibinfo {volume} {77}},\ \bibinfo {pages}
  {3865} (\bibinfo {year} {1996})}\BibitemShut {NoStop}%
\bibitem [{\citenamefont {Grimme}(2006)}]{grimme2006semiempirical}%
  \BibitemOpen
  \bibfield  {author} {\bibinfo {author} {\bibfnamefont {S.}~\bibnamefont
  {Grimme}},\ }\href
  {http://onlinelibrary.wiley.com/doi/10.1002/jcc.20495/abstract} {\bibfield
  {journal} {\bibinfo  {journal} {J. Comp. Chem.},\ }\textbf {\bibinfo {volume}
  {27}},\ \bibinfo {pages} {1787} (\bibinfo {year} {2006})}\BibitemShut
  {NoStop}%
\bibitem [{\citenamefont {Adamo}\ and\ \citenamefont
  {Barone}(1999)}]{adamo1999toward}%
  \BibitemOpen
  \bibfield  {author} {\bibinfo {author} {\bibfnamefont {C.}~\bibnamefont
  {Adamo}}\ and\ \bibinfo {author} {\bibfnamefont {V.}~\bibnamefont {Barone}},\
  }\href {http://jcp.aip.org/resource/1/jcpsa6/v110/i13/p6158_s1} {\bibfield
  {journal} {\bibinfo  {journal} {J. Chem. Phys.},\ }\textbf {\bibinfo {volume}
  {110}},\ \bibinfo {pages} {6158} (\bibinfo {year} {1999})}\BibitemShut
  {NoStop}%
\bibitem [{\citenamefont {Vanderbilt}(1990)}]{PhysRevB.41.7892}%
  \BibitemOpen
  \bibfield  {author} {\bibinfo {author} {\bibfnamefont {D.}~\bibnamefont
  {Vanderbilt}},\ }\Doi {10.1103/PhysRevB.41.7892} {\bibfield  {journal}
  {\bibinfo  {journal} {Phys. Rev. B},\ }\textbf {\bibinfo {volume} {41}},\
  \bibinfo {pages} {7892} (\bibinfo {year} {1990})}\BibitemShut {NoStop}%
\bibitem [{web(2011)}]{website:opium}%
  \BibitemOpen
  \href@noop {} {\bibfield  {journal} {\bibinfo  {journal} {Opium:
  Pseudopotential Generation Project. http://opium.sourceforge.net/}} (\bibinfo
  {year} {2011})}\BibitemShut {NoStop}%
\bibitem [{\citenamefont {Monkhorst}\ and\ \citenamefont
  {Pack}(1976)}]{PhysRevB.13.5188}%
  \BibitemOpen
  \bibfield  {author} {\bibinfo {author} {\bibfnamefont {H.~J.}\ \bibnamefont
  {Monkhorst}}\ and\ \bibinfo {author} {\bibfnamefont {J.~D.}\ \bibnamefont
  {Pack}},\ }\Doi {10.1103/PhysRevB.13.5188} {\bibfield  {journal} {\bibinfo
  {journal} {Phys. Rev. B},\ }\textbf {\bibinfo {volume} {13}},\ \bibinfo
  {pages} {5188} (\bibinfo {year} {1976})}\BibitemShut {NoStop}%
\bibitem [{\citenamefont {Fortes}(2010)}]{fortes2010report}%
  \BibitemOpen
  \bibfield  {author} {\bibinfo {author} {\bibfnamefont {A.~D.}\ \bibnamefont
  {Fortes}},\ }\href
  {http://www.homepages.ucl.ac.uk/~ucfbanf/publications/reports/5_24_423.pdf}
  {\bibfield  {journal} {\bibinfo  {journal} {Institut Laue Langevin
  experimental report},\ \bibinfo {pages} {5}} (\bibinfo {year}
  {2010})}\BibitemShut {NoStop}%
\bibitem [{\citenamefont {Fortes}\ \emph {et~al.}(2007)\citenamefont {Fortes},
  \citenamefont {Wood}, \citenamefont {Alfredsson}, \citenamefont
  {Vo\u{c}adlo}, \citenamefont {Knight}, \citenamefont {Marshall},
  \citenamefont {Tucker},\ and\ \citenamefont
  {Fernandez-Alonso}}]{doi:10.1080/08957950701265029}%
  \BibitemOpen
  \bibfield  {author} {\bibinfo {author} {\bibfnamefont {A.~D.}\ \bibnamefont
  {Fortes}}, \bibinfo {author} {\bibfnamefont {I.~G.}\ \bibnamefont {Wood}},
  \bibinfo {author} {\bibfnamefont {M.}~\bibnamefont {Alfredsson}}, \bibinfo
  {author} {\bibfnamefont {L.}~\bibnamefont {Vo\u{c}adlo}}, \bibinfo {author}
  {\bibfnamefont {K.~S.}\ \bibnamefont {Knight}}, \bibinfo {author}
  {\bibfnamefont {W.~G.}\ \bibnamefont {Marshall}}, \bibinfo {author}
  {\bibfnamefont {M.~G.}\ \bibnamefont {Tucker}}, \ and\ \bibinfo {author}
  {\bibfnamefont {F.}~\bibnamefont {Fernandez-Alonso}},\ }\href
  {http://www.tandfonline.com/doi/abs/10.1080/08957950701265029} {\bibfield
  {journal} {\bibinfo  {journal} {High Press. Res.},\ }\textbf {\bibinfo
  {volume} {27}},\ \bibinfo {pages} {201} (\bibinfo {year} {2007})}\BibitemShut
  {NoStop}%
\bibitem [{\citenamefont {Loveday}\ \emph {et~al.}(2009)\citenamefont
  {Loveday}, \citenamefont {Nelmes}, \citenamefont {Bull}, \citenamefont
  {Maynard-Casely},\ and\ \citenamefont
  {Guthrie}}]{doi:10.1080/08957950903162057}%
  \BibitemOpen
  \bibfield  {author} {\bibinfo {author} {\bibfnamefont {J.~S.}\ \bibnamefont
  {Loveday}}, \bibinfo {author} {\bibfnamefont {R.~J.}\ \bibnamefont {Nelmes}},
  \bibinfo {author} {\bibfnamefont {C.~L.}\ \bibnamefont {Bull}}, \bibinfo
  {author} {\bibfnamefont {H.~E.}\ \bibnamefont {Maynard-Casely}}, \ and\
  \bibinfo {author} {\bibfnamefont {M.}~\bibnamefont {Guthrie}},\ }\href
  {http://www.tandfonline.com/doi/abs/10.1080/08957950903162057} {\bibfield
  {journal} {\bibinfo  {journal} {High Press. Res.},\ }\textbf {\bibinfo
  {volume} {29}},\ \bibinfo {pages} {396} (\bibinfo {year} {2009})}\BibitemShut
  {NoStop}%
\bibitem [{\citenamefont {Birch}(1947)}]{birch1947finite}%
  \BibitemOpen
  \bibfield  {author} {\bibinfo {author} {\bibfnamefont {F.}~\bibnamefont
  {Birch}},\ }\href {http://prola.aps.org/abstract/PR/v71/i11/p809_1}
  {\bibfield  {journal} {\bibinfo  {journal} {Phys. Rev.},\ }\textbf {\bibinfo
  {volume} {71}},\ \bibinfo {pages} {809} (\bibinfo {year} {1947})}\BibitemShut
  {NoStop}%
\bibitem [{\citenamefont {Fortes}\ \emph {et~al.}(2011)\citenamefont {Fortes},
  \citenamefont {Griffiths}, \citenamefont {Needs}, \citenamefont {Pickard},\
  and\ \citenamefont {Hansen}}]{fortes2011iucr}%
  \BibitemOpen
  \bibfield  {author} {\bibinfo {author} {\bibfnamefont {A.~D.}\ \bibnamefont
  {Fortes}}, \bibinfo {author} {\bibfnamefont {G.~I.~G.}\ \bibnamefont
  {Griffiths}}, \bibinfo {author} {\bibfnamefont {R.~J.}\ \bibnamefont
  {Needs}}, \bibinfo {author} {\bibfnamefont {C.~J.}\ \bibnamefont {Pickard}},
  \ and\ \bibinfo {author} {\bibfnamefont {T.}~\bibnamefont {Hansen}},\
  }\href@noop {} {\bibfield  {journal} {\bibinfo  {journal} {Institut Laue
  Langevin experimental report, XXII Congress and General Assembly
  International Union of Crystallography, IUCr2011, Madrid, Spain}} (\bibinfo
  {year} {2011})}\BibitemShut {NoStop}%
\bibitem [{\citenamefont {Boys}\ and\ \citenamefont
  {Bernardi}(1970)}]{BoysB70}%
  \BibitemOpen
  \bibfield  {author} {\bibinfo {author} {\bibfnamefont {S.~F.}\ \bibnamefont
  {Boys}}\ and\ \bibinfo {author} {\bibfnamefont {F.}~\bibnamefont
  {Bernardi}},\ }\href@noop {} {\bibfield  {journal} {\bibinfo  {journal} {Mol.
  Phys.},\ }\textbf {\bibinfo {volume} {19}},\ \bibinfo {pages} {553} (\bibinfo
  {year} {1970})}\BibitemShut {NoStop}%
\bibitem [{\citenamefont {Sadlej}(1991)}]{Sadlej91}%
  \BibitemOpen
  \bibfield  {author} {\bibinfo {author} {\bibfnamefont {A.~J.}\ \bibnamefont
  {Sadlej}},\ }\href@noop {} {\bibfield  {journal} {\bibinfo  {journal} {Theor.
  Chim. Acta},\ }\textbf {\bibinfo {volume} {79}},\ \bibinfo {pages} {123}
  (\bibinfo {year} {1991})}\BibitemShut {NoStop}%
\bibitem [{\citenamefont {Bukowski}\ \emph {et~al.}(1999)\citenamefont
  {Bukowski}, \citenamefont {Sadlej}, \citenamefont {Jeziorski}, \citenamefont
  {Jankowski}, \citenamefont {Szalewicz}, \citenamefont {Kucharski},
  \citenamefont {Williams},\ and\ \citenamefont {Rice}}]{BukowskiSJJSKWR99}%
  \BibitemOpen
  \bibfield  {author} {\bibinfo {author} {\bibfnamefont {R.}~\bibnamefont
  {Bukowski}}, \bibinfo {author} {\bibfnamefont {J.}~\bibnamefont {Sadlej}},
  \bibinfo {author} {\bibfnamefont {B.}~\bibnamefont {Jeziorski}}, \bibinfo
  {author} {\bibfnamefont {P.}~\bibnamefont {Jankowski}}, \bibinfo {author}
  {\bibfnamefont {K.}~\bibnamefont {Szalewicz}}, \bibinfo {author}
  {\bibfnamefont {S.~A.}\ \bibnamefont {Kucharski}}, \bibinfo {author}
  {\bibfnamefont {H.~L.}\ \bibnamefont {Williams}}, \ and\ \bibinfo {author}
  {\bibfnamefont {B.~M.}\ \bibnamefont {Rice}},\ }\href@noop {} {\bibfield
  {journal} {\bibinfo  {journal} {J. Chem. Phys.},\ }\textbf {\bibinfo {volume}
  {110}},\ \bibinfo {pages} {3785} (\bibinfo {year} {1999})}\BibitemShut
  {NoStop}%
\bibitem [{\citenamefont {Williams}\ \emph {et~al.}(1995)\citenamefont
  {Williams}, \citenamefont {Mas}, \citenamefont {Szalewicz},\ and\
  \citenamefont {Jeziorski}}]{WilliamsMSJ95}%
  \BibitemOpen
  \bibfield  {author} {\bibinfo {author} {\bibfnamefont {H.~L.}\ \bibnamefont
  {Williams}}, \bibinfo {author} {\bibfnamefont {E.~M.}\ \bibnamefont {Mas}},
  \bibinfo {author} {\bibfnamefont {K.}~\bibnamefont {Szalewicz}}, \ and\
  \bibinfo {author} {\bibfnamefont {B.}~\bibnamefont {Jeziorski}},\ }\href@noop
  {} {\bibfield  {journal} {\bibinfo  {journal} {J. Chem. Phys.},\ }\textbf
  {\bibinfo {volume} {103}},\ \bibinfo {pages} {7374} (\bibinfo {year}
  {1995})}\BibitemShut {NoStop}%
\bibitem [{\citenamefont {Helgaker}\ \emph {et~al.}(2005)\citenamefont
  {Helgaker}, \citenamefont {Jensen}, \citenamefont {Joergensen}, \citenamefont
  {Olsen}, \citenamefont {Ruud}, \citenamefont {Aagren}, \citenamefont {Auer},
  \citenamefont {Bak}, \citenamefont {Bakken}, \citenamefont {Christiansen},
  \citenamefont {Coriani}, \citenamefont {Dahle}, \citenamefont {Dalskov},
  \citenamefont {Enevoldsen}, \citenamefont {Fernandez}, \citenamefont
  {Haettig}, \citenamefont {Hald}, \citenamefont {Halkier}, \citenamefont
  {Heiberg}, \citenamefont {Hettema}, \citenamefont {Jonsson}, \citenamefont
  {Kirpekar}, \citenamefont {Kobayashi}, \citenamefont {Koch}, \citenamefont
  {Mikkelsen}, \citenamefont {Norman}, \citenamefont {Packer}, \citenamefont
  {Pedersen}, \citenamefont {Ruden}, \citenamefont {Sanchez}, \citenamefont
  {Saue}, \citenamefont {Sauer}, \citenamefont {Schimmelpfennig}, \citenamefont
  {Sylvester-Hvid}, \citenamefont {Taylor},\ and\ \citenamefont
  {Vahtras}}]{DALTON2}%
  \BibitemOpen
  \bibfield  {author} {\bibinfo {author} {\bibfnamefont {T.}~\bibnamefont
  {Helgaker}}, \bibinfo {author} {\bibfnamefont {H.~J.~A.}\ \bibnamefont
  {Jensen}}, \bibinfo {author} {\bibfnamefont {P.}~\bibnamefont {Joergensen}},
  \bibinfo {author} {\bibfnamefont {J.}~\bibnamefont {Olsen}}, \bibinfo
  {author} {\bibfnamefont {K.}~\bibnamefont {Ruud}}, \bibinfo {author}
  {\bibfnamefont {H.}~\bibnamefont {Aagren}}, \bibinfo {author} {\bibfnamefont
  {A.}~\bibnamefont {Auer}}, \bibinfo {author} {\bibfnamefont {K.}~\bibnamefont
  {Bak}}, \bibinfo {author} {\bibfnamefont {V.}~\bibnamefont {Bakken}},
  \bibinfo {author} {\bibfnamefont {O.}~\bibnamefont {Christiansen}}, \bibinfo
  {author} {\bibfnamefont {S.}~\bibnamefont {Coriani}}, \bibinfo {author}
  {\bibfnamefont {P.}~\bibnamefont {Dahle}}, \bibinfo {author} {\bibfnamefont
  {E.~K.}\ \bibnamefont {Dalskov}}, \bibinfo {author} {\bibfnamefont
  {T.}~\bibnamefont {Enevoldsen}}, \bibinfo {author} {\bibfnamefont
  {B.}~\bibnamefont {Fernandez}}, \bibinfo {author} {\bibfnamefont
  {C.}~\bibnamefont {Haettig}}, \bibinfo {author} {\bibfnamefont
  {K.}~\bibnamefont {Hald}}, \bibinfo {author} {\bibfnamefont {A.}~\bibnamefont
  {Halkier}}, \bibinfo {author} {\bibfnamefont {H.}~\bibnamefont {Heiberg}},
  \bibinfo {author} {\bibfnamefont {H.}~\bibnamefont {Hettema}}, \bibinfo
  {author} {\bibfnamefont {D.}~\bibnamefont {Jonsson}}, \bibinfo {author}
  {\bibfnamefont {S.}~\bibnamefont {Kirpekar}}, \bibinfo {author}
  {\bibfnamefont {R.}~\bibnamefont {Kobayashi}}, \bibinfo {author}
  {\bibfnamefont {H.}~\bibnamefont {Koch}}, \bibinfo {author} {\bibfnamefont
  {K.~V.}\ \bibnamefont {Mikkelsen}}, \bibinfo {author} {\bibfnamefont
  {P.}~\bibnamefont {Norman}}, \bibinfo {author} {\bibfnamefont {M.~J.}\
  \bibnamefont {Packer}}, \bibinfo {author} {\bibfnamefont {T.~B.}\
  \bibnamefont {Pedersen}}, \bibinfo {author} {\bibfnamefont {T.~A.}\
  \bibnamefont {Ruden}}, \bibinfo {author} {\bibfnamefont {A.}~\bibnamefont
  {Sanchez}}, \bibinfo {author} {\bibfnamefont {T.}~\bibnamefont {Saue}},
  \bibinfo {author} {\bibfnamefont {S.~P.~A.}\ \bibnamefont {Sauer}}, \bibinfo
  {author} {\bibfnamefont {B.}~\bibnamefont {Schimmelpfennig}}, \bibinfo
  {author} {\bibfnamefont {K.~O.}\ \bibnamefont {Sylvester-Hvid}}, \bibinfo
  {author} {\bibfnamefont {P.~R.}\ \bibnamefont {Taylor}}, \ and\ \bibinfo
  {author} {\bibfnamefont {O.}~\bibnamefont {Vahtras}},\ }\href@noop {}
  {\enquote {\bibinfo {title} {Dalton, a molecular electronic structure
  program, release 2.0},}\ } (\bibinfo {year} {2005}),\ \bibinfo {note} {see
  http://www.kjemi.uio.no/software/dalton/dalton.html}\BibitemShut {NoStop}%
\bibitem [{\citenamefont {Misquitta}\ and\ \citenamefont
  {Stone}(2007)}]{CamCASP}%
  \BibitemOpen
  \bibfield  {author} {\bibinfo {author} {\bibfnamefont {A.~J.}\ \bibnamefont
  {Misquitta}}\ and\ \bibinfo {author} {\bibfnamefont {A.~J.}\ \bibnamefont
  {Stone}},\ }\href@noop {} {\enquote {\bibinfo {title} {{\sc CamCASP}: a
  program for studying intermolecular interactions and for the calculation of
  molecular properties in distributed form},}\ }\bibinfo {howpublished}
  {University of Cambridge} (\bibinfo {year} {2007}),\ \bibinfo {note}
  {http://www-stone.ch.cam.ac.uk/programs.html\#CamCASP}\BibitemShut {NoStop}%
\bibitem [{\citenamefont {Misquitta}\ and\ \citenamefont
  {Szalewicz}(2002)}]{MisquittaS02}%
  \BibitemOpen
  \bibfield  {author} {\bibinfo {author} {\bibfnamefont {A.~J.}\ \bibnamefont
  {Misquitta}}\ and\ \bibinfo {author} {\bibfnamefont {K.}~\bibnamefont
  {Szalewicz}},\ }\href@noop {} {\bibfield  {journal} {\bibinfo  {journal}
  {Chem. Phys. Lett.},\ }\textbf {\bibinfo {volume} {357}},\ \bibinfo {pages}
  {301} (\bibinfo {year} {2002})}\BibitemShut {NoStop}%
\bibitem [{\citenamefont {Misquitta}\ \emph {et~al.}(2003)\citenamefont
  {Misquitta}, \citenamefont {Jeziorski},\ and\ \citenamefont
  {Szalewicz}}]{MisquittaJS03}%
  \BibitemOpen
  \bibfield  {author} {\bibinfo {author} {\bibfnamefont {A.~J.}\ \bibnamefont
  {Misquitta}}, \bibinfo {author} {\bibfnamefont {B.}~\bibnamefont
  {Jeziorski}}, \ and\ \bibinfo {author} {\bibfnamefont {K.}~\bibnamefont
  {Szalewicz}},\ }\href@noop {} {\bibfield  {journal} {\bibinfo  {journal}
  {Phys. Rev. Lett.},\ }\textbf {\bibinfo {volume} {91}},\ \bibinfo {pages}
  {33201} (\bibinfo {year} {2003})}\BibitemShut {NoStop}%
\bibitem [{\citenamefont {Misquitta}\ \emph {et~al.}(2005)\citenamefont
  {Misquitta}, \citenamefont {Podeszwa}, \citenamefont {Jeziorski},\ and\
  \citenamefont {Szalewicz}}]{MisquittaPJS05b}%
  \BibitemOpen
  \bibfield  {author} {\bibinfo {author} {\bibfnamefont {A.~J.}\ \bibnamefont
  {Misquitta}}, \bibinfo {author} {\bibfnamefont {R.}~\bibnamefont {Podeszwa}},
  \bibinfo {author} {\bibfnamefont {B.}~\bibnamefont {Jeziorski}}, \ and\
  \bibinfo {author} {\bibfnamefont {K.}~\bibnamefont {Szalewicz}},\ }\href@noop
  {} {\bibfield  {journal} {\bibinfo  {journal} {J. Chem. Phys.},\ }\textbf
  {\bibinfo {volume} {123}},\ \bibinfo {pages} {214103} (\bibinfo {year}
  {2005})}\BibitemShut {NoStop}%
\bibitem [{\citenamefont {Cohen}\ \emph {et~al.}(2008)\citenamefont {Cohen},
  \citenamefont {Mori-S\'{a}nchez},\ and\ \citenamefont {Yang}}]{CohenM-SY08a}%
  \BibitemOpen
  \bibfield  {author} {\bibinfo {author} {\bibfnamefont {A.~J.}\ \bibnamefont
  {Cohen}}, \bibinfo {author} {\bibfnamefont {P.}~\bibnamefont
  {Mori-S\'{a}nchez}}, \ and\ \bibinfo {author} {\bibfnamefont
  {W.}~\bibnamefont {Yang}},\ }\Doi {10.1126/science.1158722} {\bibfield
  {journal} {\bibinfo  {journal} {Science},\ }\textbf {\bibinfo {volume}
  {321}},\ \bibinfo {pages} {792} (\bibinfo {year} {2008})}\BibitemShut
  {NoStop}%
\bibitem [{\citenamefont {Morris}\ and\ \citenamefont
  {Pickard}(2011)}]{lindos}%
  \BibitemOpen
  \bibfield  {author} {\bibinfo {author} {\bibfnamefont {A.~J.}\ \bibnamefont
  {Morris}}\ and\ \bibinfo {author} {\bibfnamefont {C.~J.}\ \bibnamefont
  {Pickard}},\ }\href@noop {} {\bibfield  {journal} {\bibinfo  {journal}
  {LINDOS - Version 1.3 User Manual, University College London}} (\bibinfo
  {year} {2011})}\BibitemShut {NoStop}%
\end{thebibliography}


\end{document}